\documentclass[letterpaper,11pt,twoside]{article}

\usepackage{ifpdf}
\usepackage{microtype}
\usepackage{times}
\usepackage{graphicx}
\usepackage[table]{xcolor}
\usepackage{amsmath,amsfonts}
\usepackage[hyphens]{url}
\usepackage[numbers,sort]{natbib}
\usepackage{dcolumn,booktabs}
\usepackage{multirow}
\usepackage[compact]{titlesec}
\usepackage{fancyhdr}
\usepackage{shortcut}
\usepackage{wrapfig}
\usepackage{algorithm,algorithmic,rotating}
\usepackage{paralist}
\usepackage{xspace}
\usepackage{pifont}
\usepackage{subfigure}
\usepackage{url}
\usepackage{pgfgantt}
\usepackage{pgfplots}
\usepackage{tikz}

\usepackage{hyperref}

 \hypersetup{
    colorlinks = true,
    linkcolor = red,
    anchorcolor = blue,
    citecolor = blue,
    filecolor = blue,
    urlcolor = blue
}

\pdfoutput=1

\renewcommand{\sysname}{{\sc Acoustic-Turf}\xspace}

\usepackage{color}

\ifpdf
  \pdfpagewidth=\paperwidth
  \pdfpageheight=\paperheight
\fi

\advance\textheight-\topmargin
\advance\textheight-\headheight
\advance\textheight-\headsep
\advance\textheight-\footskip
\advance\textheight12pt

\SetExpansion[context=sloppy,stretch=60,shrink=60,step=5]{encoding=*}{}

\newcolumntype{d}[1]{D{.}{.}{#1}}
\newcolumntype{R}[1]{D{/}{/}{#1}}

\usepackage{refcount}

\pgfplotsset{compat=1.13}
\begin{document}

\thispagestyle{empty}

\noindent{}

\begin{center}

\vspace{0.4in}

		\huge
        {\bf TECHNICAL REPORT}

        \vspace{0.8cm}

{\LARGE \bf 
\sysname: Acoustic-based Privacy-Preserving COVID-19 Contact Tracing 
}
\vfil

\normalsize
Yuxiang Luo, Cheng Zhang, Yunqi Zhang, Chaoshun Zuo, \\ Dong Xuan, Zhiqiang Lin, Adam C. Champion, and Ness Shroff \\
Department of Computer Science and Engineering\\
{The Ohio State University}\par

\vfil

\end{center} 
\noindent 

\begin{abstract}
	In this paper, we propose a new privacy-preserving, automated contact tracing system, \sysname, to fight COVID-19 using acoustic signals sent from ubiquitous mobile devices.
	At a high level, \sysname adaptively broadcasts inaudible ultrasonic signals with randomly generated IDs in the vicinity. Simultaneously, the system receives other ultrasonic signals sent from nearby (e.g., 6 feet) users. In such a system, individual user IDs are not disclosed to others and the system can accurately detect ``encounters'' in physical proximity with 6-foot granularity. We have implemented a prototype of \sysname on Android and evaluated its performance in terms of acoustic-signal-based encounter detection accuracy and power consumption at different ranges and under various occlusion scenarios. Experimental results show that \sysname can detect multiple contacts within a 6-foot range for mobile phones placed in pockets and outside pockets. 
	Furthermore, our acoustic-signal-based system achieves greater precision than wireless-signal-based approaches when contact tracing is performed through walls. \sysname correctly determines that people on opposite sides of a wall are \emph{not} in contact with one another, whereas the Bluetooth-based approaches detect nonexistent contacts among them. \looseness=-1
\end{abstract}

\par
 
\vfil

\clearpage

\advance\baselineskip0pt plus.5pt minus0pt
\flushbottom

\pagenumbering{arabic}
\setcounter{page}{1}

\section{Introduction}

The outbreak of the novel coronavirus (COVID-19) has unfolded as an unprecedented worldwide crisis with enormous social and economic impacts. In many countries today, public health authorities have mandated \emph{social distancing} (i.e., 6-foot separation, according to the CDC). Accordingly, travel has been restricted, schools have been shut down, shops have been closed,  shelter-in-place orders have been issued, and remote work has become the norm. While vaccines or therapeutic drugs can fundamentally stop the COVID-19 pandemic, it is unlikely that they will reach the market soon (estimates range between 12 and 18 months away~\cite{Expertsw13:online}). As such, we have to live with the virus in the interim, which necessitates techniques and tools for containment that also enable safe reopening of society and the economy. \looseness=-1

At a high level, in addition to social distancing, practical approaches to contain the spread of the pandemic include ($i$) large-scale testing; and ($ii$) aggressive contact tracing, as countries such as South Korea~\cite{Coronavi89:online} have demonstrated. Large-scale testing can quickly determine who has been infected and who has recovered, applying different strategies to respective groups; yet we must rely on medical science to solve the testing problem. Aggressive contact tracing can quickly identify those in the former group and recommend immediate quarantine within a certain period of time (e.g., two weeks) after a positive virus test result. However, manually performing such tracking would be impossible given large populations who may have been exposed to the virus. Hence, we need automated contact tracking systems. \looseness=-1

In the past few months, researchers have made significant progress developing automated contact tracing systems using various location data generated from cameras, cellular towers, credit-card transactions, GPS signals, Wi-Fi hotspots, and Bluetooth signals. An effective automated contact tracing system must preserve the privacy of users; otherwise it risks becoming an invasive surveillance system. As Bluetooth achieves greater privacy and accuracy than alternatives such as GPS, it is widely considered that the use of Bluetooth signals is a practical approach~\cite{Explaine84:online}. Built atop Bluetooth, numerous privacy-preserving protocols including CTV~\cite{Canetti2020AnonymousCD}, East-PACT~\cite{PACT},  West-PACT~\cite{Chan2020PACTPS}, DP3T~\cite{DP3T}, COVID-Watch~\cite{CovidWatch}, and TCN~\cite{TCN} have been developed, and open source tools such as TraceTogether~\cite{TraceTogether} and CovidSafe~\cite{CovidSaf66} have been released. To facilitate tracing app development and address other technical challenges (such as limited battery life), Apple and Google have jointly developed Privacy-Preserving Contact Tracing in the iOS and Android operating systems and provided corresponding API support~\cite{AG:PPCT}.

While Bluetooth-based automatic tracking has shown great promise in fighting against COVID-19, it still faces two major challenges: large-scale user adoption and proximity accuracy. The efficacy of a tracking system depends on its use by $\sim$50--70\% of the population~\cite{TheApple44:online,CovidWatch}. Designing a privacy-preserving, battery-saving system could help attract more users. However, achieving accurate proximity measurements via signal strength remains a challenging problem~\cite{Bluetoot:Challenges}. Specifically,
Bluetooth's received signal strength indicator (RSSI) is used to measure the distance between two phones. In theory, stronger signals indicate closer phones. However, in reality, signal strength is affected by various factors such as phone orientation (i.e., portrait vs. landscape mode), surrounding objects and surfaces (e.g., indoors vs. outdoors, pockets vs. purses), and even phone operating systems (i.e., iPhone vs. Android)~\cite{Bluetoot:Challenges}.

\begin{figure}[t]
	\centering
	\includegraphics[width=1\linewidth]{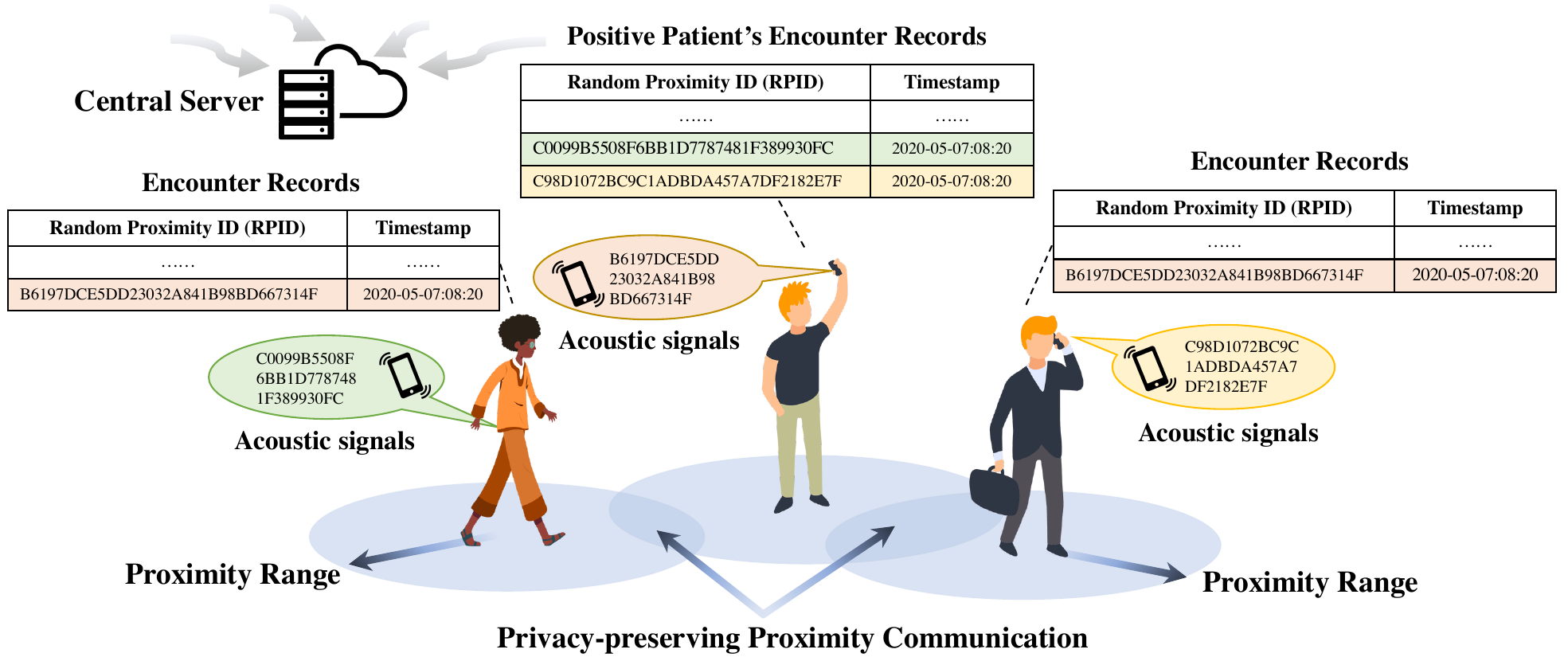}
	\caption{Privacy preserving contact tracing of \sysname.}
	\label{fig:scenario}
\end{figure}

Therefore, integrating other mobile phone sensors is imperative in order to measure proximity precisely. In this paper, we describe how to leverage preexisting ultrasonic sensors in phones for automated contract tracing.
Compared to Wi-Fi or Bluetooth signals, which are typically determined by hardware 
and influenced by the nearby environment, acoustic signals grant us a unique advantage: software can have a completely control over their communication. Based on this observation, we present \sysname, a new automated contact tracing system using acoustic signals from ubiquitous mobile devices without any additional hardware support. Similar to many other privacy-preserving systems, \sysname provides a high degree of privacy in the following ways: ($i$) The communication (sensing) range of acoustic signals on mobile phones is naturally limited compared with those of Wi-Fi and Bluetooth signals; and ($ii$) each user's hardware IDs (such as Wi-Fi and Bluetooth MAC addresses) are neither used nor disclosed to nearby contacts. Instead, only randomly-generated IDs are disclosed to contacts, which guards against potential trajectory-based attacks.

An illustration of \sysname in action is provided in
\autoref{fig:scenario}. One salient feature of \sysname is that it detects contacts among mobile phone users in physical proximity.  
Specifically, assume there are users Alice and Bob. \sysname generates Alice's random ID and sends it to nearby users via acoustic signals. Bob, who is nearby, receives her random ID and stores it locally on his phone along with the timestamp; we term this a \emph{contact record}. Later, Bob is diagnosed with COVID-19. He uploads his contact records of his choice to a central server maintained by public health authorities (such as the CDC). If Alice contracts the virus, the server can trace her contacts via Bob's uploaded contact records and notify her of the infection of a person with whom she had contact. In \sysname, these records do not reveal any Alice and Bob's identities; \sysname only reveals the random IDs of people with whom they were in contact at the corresponding times. Due to the range of acoustic communication, a listening adversary would have to be within this range to detect the random IDs, but \sysname updates them frequently.

In summary, this paper makes the following contributions: (1) To the best of our knowledge, \sysname is the first effort using acoustic-signal-based communication and sensing for COVID-19 contact tracing. (2) We design a novel acoustic broadcast protocol for working scenarios in which communications are casual, unintentional, non-cooperative, and opportunistic among up to ten people in occluded environments with limited power consumption. (3) We implement \sysname and its acoustic broadcast protocol on commercial off-the-shelf (COTS) mobile phones running Android. (4) Further, we evaluate \sysname's performance in terms of acoustic-signal-based encounter detection accuracy and power consumption under different ranges and occlusion scenarios. Our experimental results show that \sysname can successfully detect multiple contacts within a 6-foot range for both in-pocket and out-of-pocket scenarios.

The rest of the paper is organized as follows. In \S\ref{sec:background}, we discuss related work including Bluetooth-based and ultrasonic-based proximity sensing in order to understand our techniques.  
In \S\ref{sec:overview}, we provide an overview of \sysname, followed by detailed design in \S\ref{sec:design}. \S\ref{sec:implemetnation} shares implementation details and \S\ref{sec:evaluation} present the evaluation results. We discuss limitations and future work in \S\ref{sec:discussion}, and we conclude in \S\ref{sec:conclusion}.

\section{Related Work}
\label{sec:background}

\vspace{-0.1in}
\paragraph{Automated Privacy-Preserving Mobile Contact Tracing} Recently, numerous automated privacy-preserving contact tracing frameworks have been proposed. As summarized in~\autoref{tab:summary}, these frameworks use cryptographic techniques to generate various random identifiers (IDs), which they broadcast to nearby users via Bluetooth technology. One key approach to achieve privacy preservation is by periodically changing these random IDs such that an adversary cannot link them to specific users.

\begin{table}        																			
\centering\resizebox{1\textwidth}{!}{																			
\begin{tabular}{ lccc cccc}																			
																			
		&	 \multicolumn{4}{c}{\bf Random Crypto Secrets Generation			}				&	\multicolumn{2}{c}{\bf Tracing}			&	 \textbf{Reporting}	\\	\cmidrule(rl){2-5} \cmidrule(rl){6-7} \cmidrule(rl){8-8}		
	\textbf{Framework}	&	At	&	Random	&		&	Update	&	\texttt{setTxPower}	&	Proximity 	&	Reported	\\			
		&	 Client?	&	Secret (IDs)	&	\# bits	&	Frequency	&	\texttt{Level}	&	Measurement	&	Item	\\\toprule			
																			
	CTV~\cite{Canetti2020AnonymousCD}	&	\tick	&	Token	&	N/A	&	1 Minute	&	N/A	&	N/A	&	Token	\\\hline			
																			
	\multirow{2}{*}{DP3T~\cite{DP3T}}	&	\multirow{2}{*}{\tick}	&	SK$_t$	&	256	&	24 Hour	&	\multirow{2}{*}{N/A}	&	\multirow{2}{*}{Threshold}	&	 \multirow{2}{*}{SK$_t$, t}	\\ 			
		&		&	EphIDs	&	128	&	1 Minute	&		&		&		\\\hline			
																			
	\multirow{2}{*}{East-PACT~\cite{PACT}}	&	\multirow{2}{*}{\tick}	&	Seed	&	256	&	1 Hour	&	\multirow{2}{*}{N/A}	&	\multirow{2}{*}{N/A}	&	 	\\ 			
		&		&	Chirp	&	224	&	$n$ Minutes	&		&		&	Seed, t	\\\hline			
																			
		&		&	S$_0$	&	128	&	Infection Period	&		&		&	 	\\ 			
	West-PACT~\cite{Chan2020PACTPS}	&	\tick	&	S$_i$	&	128	&	$n$ Hour	&	MEDIUM	&	Threshold	&	S$_i$, t	\\			
		&		&	ID$_i$	&	128	&	$n$ Minute	&		&		&	 	\\ \hline			
																			
		&		&	RAK, RVK	&	256	&	Infection Period	&		&	Customized	&		\\			
	TCN~\cite{TCN}	&	\tick	&	TCK	&	256	&	$n$ Hour	&	MEDIUM	&	Algorithm	&	RVK, TCK, t	\\\			
		&		&	TCN	&	128	&	15 Minute	&		&		&		\\\hline			
																			
		&		&	TEK	&	128	&	24 Hour	&		&		&		\\			
	Apple \& Google~\cite{AG:PPCT}	&	\tick	&	RPIK	&	128	&	24 Hour	&	N/A	&	N/A	&	TEK, t	\\			
		&		&	RPI	&	128	&	15 Minute	&		&		&		\\\hline			
																			
	\multirow{2}{*}{ROBERT~\cite{ROBERTRO82:online}}	&	\multirow{2}{*}{\tickNo}	&	ID	&	40	&	Infection Period	&	\multirow{2}{*}{N/A}	&	\multirow{2}{*}{N/A}	&	\multirow{2}{*}{EBID, $t$}	\\			
		&		&	EBID	&	64	&	15 Minutes	&		&		&		\\\hline			
																			
	TraceTogether~\cite{TraceTogether}	&	\tickNo	&	TempID	&	672	&	15 Minutes	&	HIGH	&	Post-Processing	&	TempIDs	\\\hline			
																			
		&		&	Seed	&	256	&	Infection Period	&		&		&		\\			
	\sysname	&	\tick	&	RDID	&	128	&	24 Hour	&	-	&	-	&	RDID, $t$	\\ 			
		&		&	RPID	&	128	&	1 Minute	&		&		&	 	\\ 			
																\bottomrule			
\end{tabular}																			
}																			
\caption{Comparison of the recently proposed automated contact tracing frameworks. EphID: Ephemeral Identifier; SK: Secret Key; TCN: Temporary Contact Number; RAK: Report Authorization Key; TCK: Temporary Contact Key; RVK: Report Verification Key; TEK: Temporary Exposure Key; RPIK: Rolling
Proximity Identifier Key; EBID: Ephemeral Bluetooth Identifier.}																			
\label{tab:summary}																			
\end{table}

In particular, there are two types of approaches to generate random IDs: generating them on phones or on servers. Most frameworks generate random IDs on phones, except ROBERT~\cite{ROBERTRO82:online} and TraceTogether~\cite{TraceTogether}, in which IDs are generated on servers. This leads to two different detection approaches: \emph{decentralized} ones, in which each user's phone determines whether the user's proximity to an infected individual, and \emph{centralized} ones, in which the central server determines user proximity. Most frameworks are decentralized. 

In addition, frameworks use different names for random IDs and users generate these IDs using different cryptographic algorithms (e.g., hash chains in DP3T~\cite{DP3T} and TCN~\cite{TCN}, and HMACs in Apple and Google~\cite{AG:PPCT}). 
Random IDs are updated at different intervals. Except initial seeds (e.g., $S_0$ in West-PACT~\cite{Chan2020PACTPS}) that never change during the entire infection period, the others either update when the Bluetooth MAC address changes (e.g., RPI, TCN, and EBID), or are updated daily (e.g., TEK, RDID). In decentralized approaches, when infected users receive positive diagnoses, these users' random IDs are uploaded to the central server with corresponding timestamps, and these IDs are used to regenerate the broadcast IDs if the latter derive from the former. In contrast, centralized approaches upload contacted users' random IDs to the central server.
 
To determine if a nearby user is a close contact, all of the existing frameworks use Bluetooth technology, which many of our computing devices (including laptops, smartphones, and wearables) employ for short-range wireless communications. 
To measure proximity between two Bluetooth Low Energy (BLE) devices---a central one (such as a smartphone) and a peripheral one (such as another smartphone)---there are two fundamental sources of information:

\begin{itemize}
    \item \textbf{Received signal strength indicator (RSSI)}. The most intuitive approach to approximate the distance from a receiver is to use the RSSI from the signal advertised from Bluetooth beacons~\cite{zafari2019survey}. Android employs the API \texttt{getRssi()}~\cite{ScanResu12:online} by which a mobile application (app) obtains the received signal strength (in dBm), where the app measures distance based on a pre-computed threshold.
    
    \item \textbf{Transmit power (TxPower)}. A Bluetooth peripheral can specify the TxPower field in the broadcast packet to inform the receiver (e.g., the smartphone) of the BLE signal's transmit power.  Android provides the API \texttt{getTxPowerLevel()}~\cite{ScanReco97:online} for this purpose. Meanwhile, peripheral devices can use the API \texttt{setTxPowerLevel()}~\cite{Advertis93:online}  to control the transmission power level for advertising that currently supports four levels:  \texttt{ULTRA\_LOW}, \texttt{LOW}, \texttt{MEDIUM}, and \texttt{HIGH}. 
    
\end{itemize}

Therefore, when using BLE for contact tracing with smartphones, the sender smartphone can control the advertisement signal's TxPower and the receiver smartphone uses RSSI (and, optionally, TxPower) to determine a contact's proximity. Although employing both RSSI and TxPower seems appealing, this is actually a complicated process~\cite{al2017improving}. Through manual inspection of the source code of the available contact tracing apps, we have found that none of them use such an approach. In particular, as \autoref{tab:summary} shows, all available source code sets the  \texttt{setTxPowerLevel()} parameter to either \texttt{MEDIUM} or \texttt{HIGH} transmission power. To detect whether a contact is in proximity, two apps use an \texttt{RSSI\_THRESHOLD} value, and two others use custom algorithms.   

Note that TraceTogether~\cite{TraceTogether} does not directly determine whether there a contact is nearby. Instead, it uploads all metadata (including phone model and RSSI) to the back-end server, which determines proximity. Apparently, TraceTogether's developers made this decision since different phone models can yield varying RSSIs at the same distance with the same TxPower (as phones may employ different chipsets and/or antennas), according to data released on GitHub~\cite{{opentrac5:online}}. Consequently, TraceTogether performs calibration at the back-end, which indicates the complexity of the process of using Bluetooth for accurate proximity measurement.

\begin{table}[t]
  \centering
  \begin{tabular}{lllll}
  \textbf{Approach}  & \textbf{\# Devices} & \textbf{Relative Motion Support} & \textbf{Distance (m)} & \textbf{Device Position} \\\hline
    Hush~\cite{novak2018ultrasound} & 2 & \tickNo & $< 0.5$ & On obstacle\\\hline
    U-Wear~\cite{santagati2015u} & $> 2$ & \tickNo & $< 0.5$ & On body (fixed)\\\hline
    Dhwani~\cite{nandakumar2013dhwani} & 2 & \tickNo & $< 0.2$\footnotemark & No obstacles\\\hline
    DopEnc~\cite{zhang2017acoustic} & 1--3 & \tickNo & 1--1.5 & N/A \\\hline
   \sysname & Up to 10 & \tick & 2 & Any position, occluded \\\bottomrule
  \end{tabular}
  \vspace*{1mm}
  \caption{Comparison of multi-access communications using acoustic signals.}%
  \label{tab:compare_mac}
\end{table}
\footnotetext{Dhwani's short distance is for security reasons.}

\paragraph{Proximity sensing using acoustic signals}
Acoustic signals, particularly using sound frequencies higher than the audible limit of human hearing (around 20 kHz), can also be used for proximity sensing and, hence, contact tracing.  The basic principle is quite simple: the speaker in the phone sends out ultrasonic pulses and the microphone receives corresponding pulses. The ultrasonic pulses can be used to broadcast random IDs generated by cryptographic techniques, just as Bluetooth-based contact tracing frameworks use Bluetooth signals for the same purpose. Actually, modern mobile phones have all necessary sensors (speakers and microphones) available for acoustic-signal-based contact tracing. 

There have been efforts to enable acoustic communications between devices that are equipped with microphones and speakers, as summarized in~\autoref{tab:compare_mac}. In particular, Novak et al. presented Hush~\cite{novak2018ultrasound}, a software modem that utilizes very high frequency sound to send data between commodity smart mobile devices. Hush performs fast modulation unobtrusively among users and achieves a low error rate. The system incorporates a fingerprinting scheme that hinders masquerades by attackers by allowing the receiver to learn and recognize packets sent
from the intended sender. However, the communication distance spans only 5--20\,cm, which is not feasible for accurate contact tracing. U-Wear~\cite{santagati2015u} enables data dissemination between ultrasonic wearable devices. Dhwani~\cite{nandakumar2013dhwani} employs audible sound for near field communications between smartphones using OFDM and FSK modulations. This work focuses on high data rate and throughput for communications. DopEnc~\cite{zhang2017acoustic} can automatically identify persons with whom users interact and considers coordinating multiple access in
order to measure the Doppler effect. \looseness=-1

While there are efforts using acoustic signals for distance proximity by measuring the received pulses, they are inaccurate particularly in dynamic and noisy environments and cannot serve contact tracing well. In our \sysname, there is no need to measure the received pulses. Compared to Bluetooth signals, programmers have far more control with acoustic signals. Note that in~\autoref{tab:compare_mac}, we list \sysname as supporting up to 10 devices tracing within a 6-foot range; this number is based solely on the practicality of how many people can be nearby within this range. The algorithm proposed in \sysname is not limited to this number and it can support even more devices. 

Most recently, concurrent to our work, Loh et al.~\cite{NOVID41:online} proposed integrating Bluetooth and ultrasonic signals for distance-aware automated contact tracing and implemented a system called NOVID. Since our \sysname purely depends on acoustic signals for contact tracing, we wish to provide a detailed comparison with NOVID in this report. However, there is no public available technical details on how NOVID works. A detailed comparison will be provided once such details are available.

\section{Overview}
\label{sec:overview}

\begin{figure}[t]
    \centering
    \includegraphics[width=5.3in]{{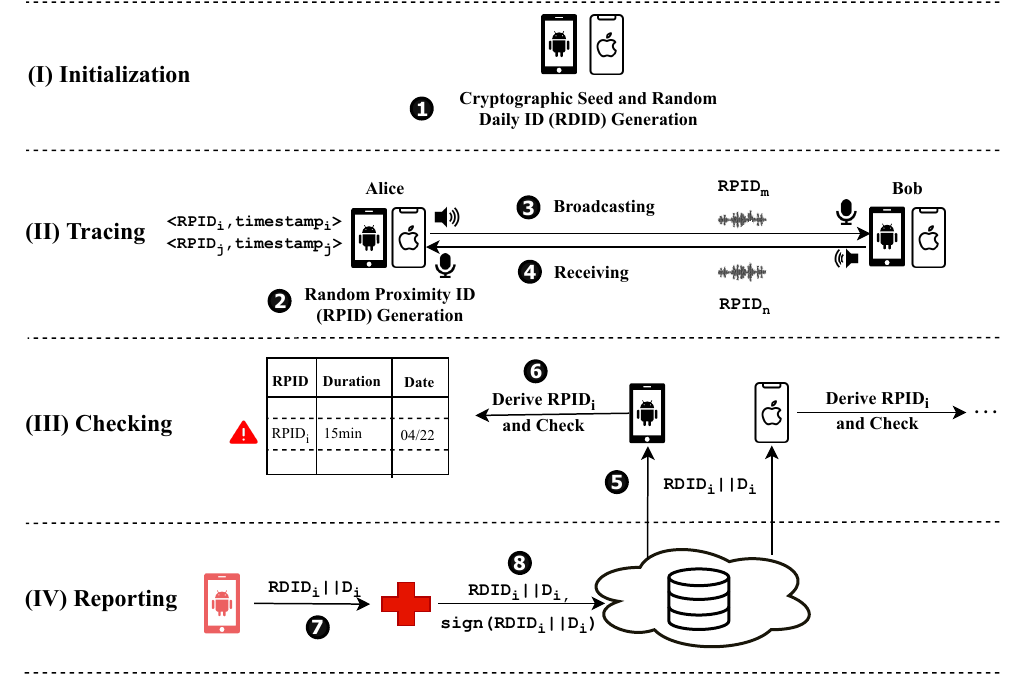}}
    \caption{Overview of \sysname.}
    \label{fig:protocol}
\end{figure}

To address the limitation of low proximity accuracy when using Bluetooth technology, we propose \sysname, an acoustic-signal-based approach for contact tracing. \sysname has similar privacy-preserving aspects as many other frameworks (e.g., ~\cite{Canetti2020AnonymousCD,PACT,Chan2020PACTPS,DP3T,TCN,CovidWatch}), but differs in how we determine proximity among contacts. \autoref{fig:protocol} presents an overview of \sysname. At a high level, \sysname has four stages (and eight steps): (I) initialization; (II) tracing; (III) checking; and (IV) reporting. We detail each stage (and correponding steps) below.

\begin{description}
    \item \textbf{(I) Initialization.} When a mobile user first participates in \sysname, she downloads our mobile tracking application (app) from the corresponding app store (e.g., Google Play or the Apple App Store) and installs it on her phone. The installation (Step \ding{182}) generates a per-device, unique, and random \emph{seed} ($s$), from which we cryptographically derive the \emph{random daily ID} ($RDID$). We derive the \emph{random proximity ID} ($RPID$) from the $RDID$. The system reports the $RDID$s to a trusted authority if the phone's owner is diagnosed with COVID-19.  \looseness=-1

          \medskip
    \item \textbf{(II) Tracing.} When the tracing app runs on a particular day (e.g., $D_i$), it picks up the corresponding $RDID_j$, from which we generate $RPID$s (Step \ding{183}). We choose to generate a random $RPID$ every minute in order to prevent nearby attackers tracking users. Next, a particular $RPID$ (e.g., $RPID_m$, 16 bytes of data) is broadcast via the phone speaker at certain frequencies to nearby phones during that particular time window (Step \ding{184}). Meanwhile, the app receives an $RPID_n$ via the microphone from nearby phones and stores the $RPID_n$ in a log on the device along with the timestamp (Step \ding{185}).

          \medskip
    \item \textbf{(III) Checking.} While the app performs contact tracing, another thread periodically fetches positive $DRID$s along with corresponding dates $D$ from the trusted authority (Step \ding{186}). From these $DRID$s, the app derives all $RPID$s associated with that particular day $D$ using the same algorithm as in the Tracing stage, and compares these $RPID$s with the stored logs on the phone (Step \ding{187}). Crucially, we need to check the date, as an attacker can generate spoofed $RPID$s after learning a $DRID$ and broadcast these $RPID$s to victims, which causes false positives. Via timestamps in the log, we can easily calculate the period of time the devices (and their owners) were in contact with each other.
          \medskip

    \item \textbf{(IV) Reporting.} Whenever a user is diagnosed with COVID-19, all of the user's $DRID$s and $D$s for the past 14 days are uploaded to a trusted authority's server (Step \ding{188}). To ensure data authenticity, we employ a healthcare authority to verify this information and sign it (Step \ding{189}), such that adversaries cannot introduce false positives (e.g., by uploading their own $RDID$s) to mislead other \sysname users.

\end{description}

\section{Detailed Design}
\label{sec:design}

In this section, we present the detailed design of \sysname. We first present how to use cryptograhic function to generate the random IDs used in our system in \S\ref{sub:protocol}, and then describe the acoustic random ID broadcasting casting protocol in \S\ref{sub:mechanism}.

\subsection{Cryptographic Random ID Generation}
\label{sub:protocol}

\vspace{-0.1in}
\paragraph{Cryptographic Seed Generation} To make \sysname simpler, all of the random IDs are derived from one  \emph{seed} ($s$), which is a per-device, unique, random number generated at install time. In our design, $s$ is 32 bytes long, and generated by cryptographic \emph{pseudorandom number generator} available in mobile operating systems such as Android. It never leaves the phone, and can be stored in a secure area such as TrustZone if that is available. It should also not be read by any malware.

\begin{wrapfigure}{r}{0.35\textwidth}
    
    \centering
    \includegraphics[width=2.7in]{{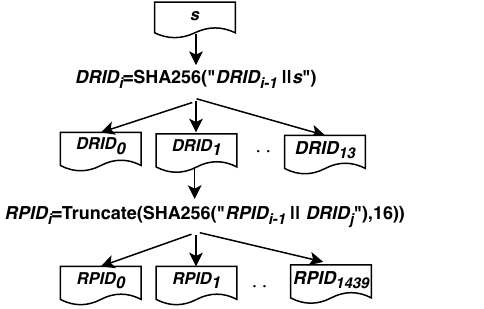}}
    
    \caption{Cryptographic ID generation in \sysname.}
    \label{fig:idg}
\end{wrapfigure}

\paragraph{Daily Random Identifier Generation} With the random \emph{seed} ($s$), we can generate a large number of daily random identifiers ($DRID$s). In our design, we use as a hash function: 
\begin{displaymath}
    DRID_i=SHA256("DRID_{i-1} || s"),
\end{displaymath}
\noindent where $s$ is our initial random \textit{seed}, and the $DRID_{-1}$ is padded with 32 bytes of \texttt{0}s. We can generate a set of $DRID$s over a period of time (e.g., 14 days), as, currently, it is considered sufficient to only track the encounters over this time, though it can be changed as needed. The $DRID$s (each of which is 32 bytes) are used by the application (app) each day to generate the random proximity identifiers ($RPID$s), which are discussed next. The app can only access some $DRID$s  when they are uploaded to trusted authorities (if the user tests positive for COVID-19). \looseness=-1

\paragraph{Random Proximity Identifier  Generation} When the tracing app runs on a particular day (e.g., $D_j$), it picks up a $DRID_j$ for that day and generates $24*60=1,440$ random proximity identifiers ($RPID$s) in Step \ding{183}. To save bandwidth, we choose to use 16-byte RPIDs, each of which is also generated vai a hash function \looseness=-1
\begin{displaymath}
    RPID_i=Truncate(SHA256(``RPID_{i-1} || DRID_j"),16)),
\end{displaymath}
where $RPID_{-1}$ is padded with 16 bytes of \texttt{0}s. Each generated $RPID_i$ is broadcast at its particular time interval. Meanwhile, if a person is diagnosed with COVID-19, his/her $DRID$ is uploaded to the central server. All other participants in the system will retrieve them from the server, re-generate the $RPID$s based on the corresponding $DRID$s, and check whether they were in close contact to the infected user.

\subsection{Acoustic Random ID Broadcasting}
\label{sub:mechanism}

As discussed in the previous subsection, each generated $RPID$ is broadcast at its particular time interval. According to the CDC, users who remain within 6 feet of each other for more than 10 minutes face high risks of coronavirus infection \cite{Chan2020PACTPS,Maccari2020DoWN}.
Hence, we must ensure that users receive at least one $RPID$ from one another if they have been in physical proximity for this amount of time, which is the basic requirement for our acoustic-signal-based random ID broadcast protocol.  The time interval for $RPID$ broadcast is important too; clearly, it is under 10 minutes. With smaller time intervals, more $RPID$s are broadcast, which increases reliability at the cost of battery consumption. Users set the interval at their discretion; the default is one minute.

To meet the above basic requirements, our protocol must address several key challenges. It must realize acoustic communications among mobile phones up to 6 feet away in environments that may be occluded or noisy (e.g., phones in purses or pockets). This is hard due to mobile phones' limited speaker volumes and the nature of acoustic signal propagation. Also, our protocol must support various numbers of nearby users who need to exchange $RPID$s. Finally, we must conserve energy as phones have limited battery capacity.

Yet contact tracing, by its nature, affords us several opportunities. Communications among users are not interactive. The purpose of communications is to determine the existence of contacts and broadcast a user's existence by exchanging $RPID$s. Low data rates are acceptable and each user has a fixed-length (i.e., 16-byte) $RPID$ to broadcast. As we describe below, we leverage
these opportunities in the design of our acoustic-signal-based random ID broadcast protocol. \looseness=-1

\subsubsection{The Broadcast Protocol}
\label{sec:phy-design}

\begin{figure}[t]
    \centering
    \includegraphics[width=1\linewidth]{{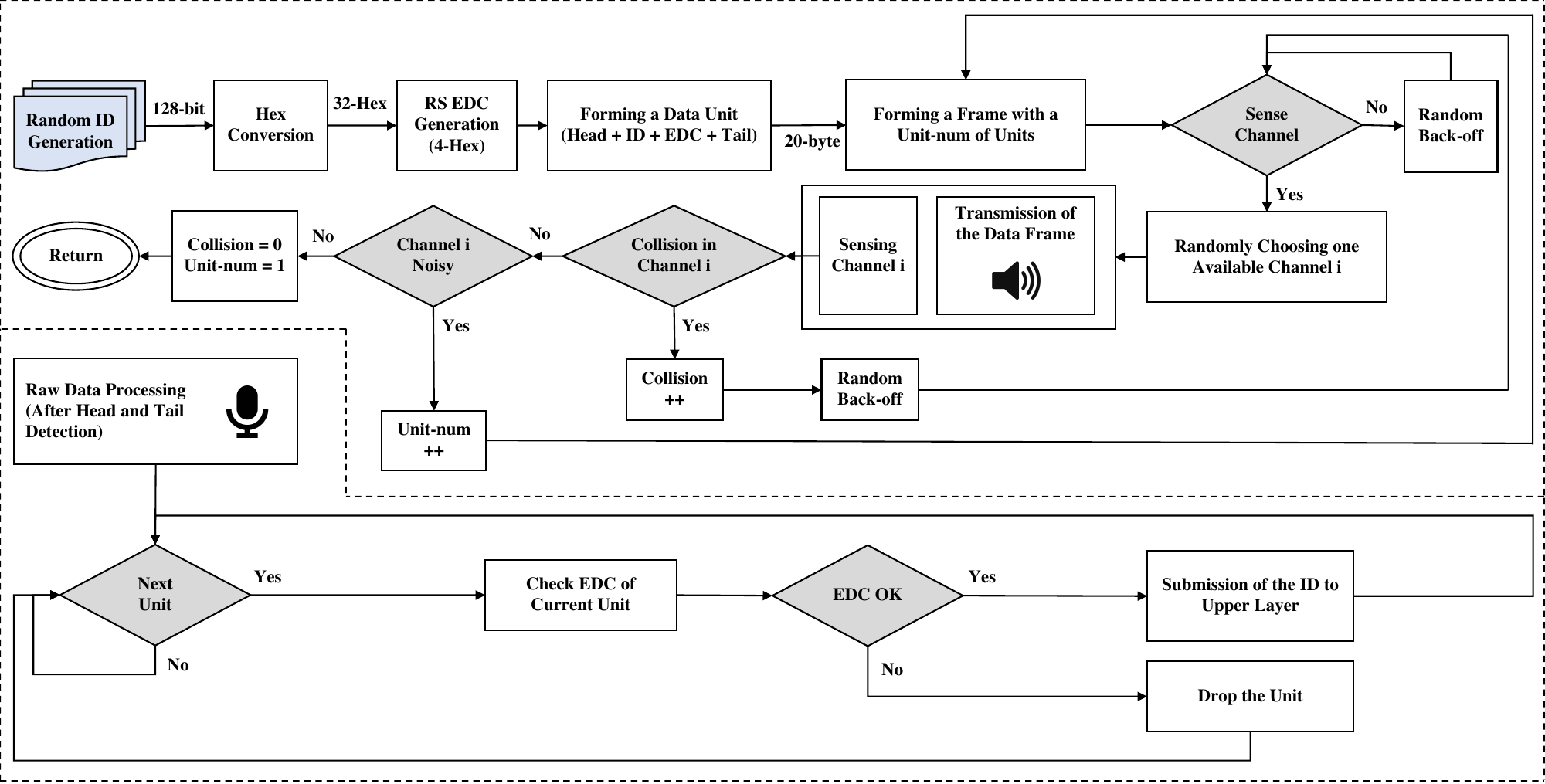}}
    \caption{The random ID broadcasting protocol.}
    \label{fig:broadcast_protocal}
\end{figure}

In the subsection, we describe our random ID broadcast protocol, which is an acoustic-channel multiple access control protocol. Our protocol adapts to changing channel and frequency occupancy, environmental noise, and distance among users by adjusting its communication channel, frequency, and reliability parameters.
\autoref{fig:broadcast_protocal} illustrates the protocol's workflow. In the following, we explain how the broadcast protocol works at the sender side and at the receiver side.

\paragraph{Sender Side} The data payload is a 128-bit $RPID$. First, the sender converts it to hexadecimal format using Reed-Solomon (RS) codes for error correction. Next, we encapsulate the $RPID$ with its corresponding RS code, a \emph{head} symbol (\texttt{H}), and a \emph{tail} symbol (\texttt{J}). \footnote{We describe the latter two symbols in the next section. We use these symbols to distinguish message boundaries.} This is the basic data unit for transmission. The sender constructs a data frame by assembling one or multiple data units based on the number of units. After the data frame is ready, the sender scans channels for availability. If one or more channels are free, the sender chooses one at random and calls the data frame transmission module, which broadcasts the frame. Meanwhile, the sender senses collisions and noise on the occupied channel. If a collision occurs, the sender abandons the transmission, updates the number of collisions, backs off a random time based on this number, and scans again for available channels. If the channel is noisy, the sender abandons the transmission, updates the unit number, and forms a new data frame based on this number. Otherwise, the sender considers the data frame as successfully broadcasted.

\paragraph{Receiver Side} First, a receiver receives a data frame by calling the data frame reception module. The receiver checks each unit received from the module. If no errors are found, the receiver retrieves the $RPID$ from the data frame; otherwise the receiver checks the next unit in the frame.

\paragraph{Remarks} In our broadcast protocol, channels are dynamically selected based on occupancy. After the sender chooses a channel, the sender  keeps sensing collisions and noise on the channel and takes action accordingly in order to avoid further collisions and achieve reliability without consuming too much power. Furthermore, the sender adjusts the number of units in a frame based on channel noise. The data unit is an important concept in the protocol (essentially, the unit contains an $RPID$ and an error detection code). At the receiver side, as long as one unit is correct, the receiver obtains the sender's $RPID$. Clearly, more units yield greater reliability. Adding units to frames does not require prior synchronization between the sender and the receiver, which is the benefit of our protocol design.

\subsubsection{Acoustic Data Frame Transmission and Reception}\label{sub_data_frame_chunk}

\begin{figure}[t]
    \centering
    \includegraphics[width=1\linewidth]{{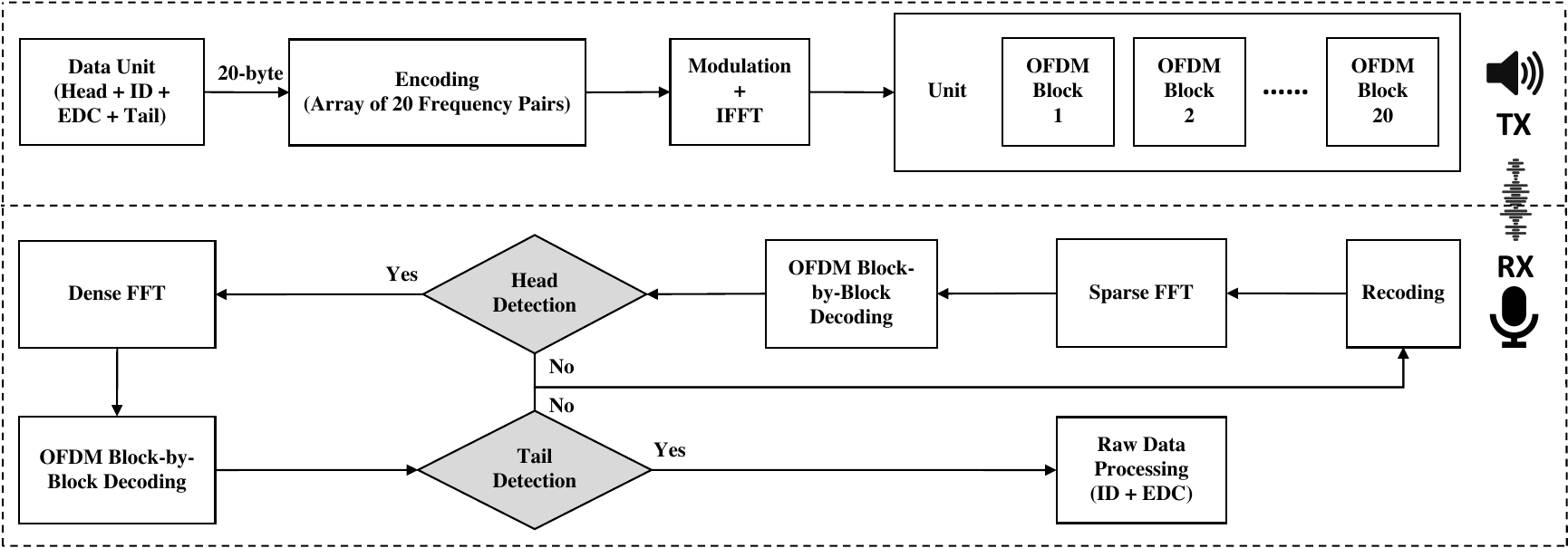}}
    \caption{Acoustic data frame transmission and reception.}
    \label{fig:comm_protocal}
\end{figure}

In the above broadcast protocol, the sender calls the data frame transmission module to broadcast the data frame using acoustic signals, and the receiver calls the data frame reception module to receive a frame. We discuss these two modules, of which \autoref{fig:comm_protocal} illustrates via flowchart. First, we describe our communications terminology.

\paragraph{Communications Terminology}
The transmission module sends a frame with one or multiple data units on an acoustic channel. Each data unit comprises 20 bytes (i.e., 160 bits). We encode data using 4-bit symbols, where each symbol is a number, a head symbol (\texttt{H}), or a tail symbol (\texttt{J}). In our protocol, we send symbols over long periods of time for reliable long-range communication. We send data symbols in frames employing binary phase-shift keying (BPSK) modulation. However, sending only one symbol at a time via the carrier wave would limit network throughput. To increase it, we send symbols in \emph{chunks}, packets of data containing two symbols sent in parallel. Our protocol supports multiple channels in total, where each user is allocated one channel at a time. To reduce interference, each channel is a combination of two different frequencies. Next, we discuss frame transmission and reception.

\paragraph{Data Frame Transmission} One data frame consists of one or multiple data units, each of which contains 160 bits (i.e., 20 bytes). In our coding mechanism, we organize each sequence of consecutive 4 bits as one symbol; hence, there are 16 different symbols. We modulate each symbol at a specific frequency using an Inverse Fast Fourier Transform (IFFT). Frequency selection follows that of OFDM. For example, when we send the symbol \texttt{0111} at frequency $f$, we modulate this symbol at the carrier wave frequency $f$. We place two symbols in a chunk; we modulate each symbol at a different frequency. The first and second symbols in a chunk correspond to an odd-numbered bit (modulated at a lower frequency) and an even-numbered bit (modulated at a higher frequency), respectively. Given a 160-bit data unit, we convert it to 40 4-bit symbols, which we place in 20 chunks. For each
symbol, we select its frequency based on a hash table mapping symbols to frequencies. We delimit the sequence of symbols using the head symbol (\texttt{H}) and the tail symbol (\texttt{J}), respectively. We create a
frequency-domain representation of the signal with $2^{15}$ (32,768) bits corresponding to the odd and even symbols. We perform an IFFT on this data to generate $2^{15}$ values using amplitude modulation (AM) in the time domain with a 48\,kHz sampling rate. We truncate the generated audio signal using the symbol duration and emit this signal via the mobile phone's speaker.

\paragraph{Data Frame Reception}
We receive an AM signal sampled at 48\,kHz, which we process using a sparse Fast Fourier Transform (FFT) for rapid symbol detection. We perform another dense FFT upon detecting a head symbol. Specifically, we receive 256 samples in an array; there are many such arrays in sequence. Each time, we perform an FFT on 2,048 samples (a 42.6\,ms audio signal). We employ a sliding window protocol with the symbol duration to scan the amplitudes based on the frequencies we selected for OFDM block-by-block decoding, If the amplitude at a given frequency exceeds a certain threshold, we regard it as a \texttt{1}. To avoid interference, we scan the low-end and high-end frequencies separately. We start and stop collecting data upon receiving head and tail symbols, respectively. We place all received symbols in a frame, which we return to the mobile app at the receiver side.

\paragraph{Remarks} The above frame transmission and reception modules provide acoustic transmission service for a given data frame. To achieve reliable communication, we adopt several mechanisms, including long symbol duration, BPSK, OFDM frequencies, and parallel frequency use (at the cost of low data rate).

\section{Implementation}
\label{sec:implemetnation}
We have developed a prototype of \sysname on Android. Our prototype consists of two major components: an \sysname mobile application (app) for each participant, and an \sysname server for public health authorities (e.g., the CDC). The server uses a database that securely stores all relevant information.
We implement the acoustic random ID boradcast protocol at the mobile app side and the security protocol between the mobile app and the server.

\paragraph{The Mobile Apps} The \sysname mobile app is a downloadable
app that periodically generates $RPID$s, broadcasts them, and
receives them from other devices. The app consists of
four fragments: \textsf{Home}, \textsf{Your Contacts}, \textsf{Your Account}, and \textsf{Settings}.\footnote{In Android development, a fragment is a composable piece of the UI that roughly corresponds to a single user-visible screen.} The \textsf{Settings} fragment lets users modify personal information and adjust app settings. The \textsf{Your Contacts} fragment contains $RPID$s received from nearby users during the last 14 days. The \textsf{Home} fragment has three pages: a \textsf{Welcome Page}, a \textsf{User Portal}, and a \textsf{Notification Center}. The \textsf{Welcome Page} briefly introduces users to \sysname and instructs them on how to use the app.
Users can login or register though the \textsf{User Portal}, which synchronizes their account credentials with the \sysname database. In the \textsf{Notification Center}, users can view notifications and warnings sent from the server. 
\looseness=-1

We generate the cryptographic \emph{seed} via the Java class \texttt{java.util.Random}; we save \emph{seed} via \texttt{SharedPreferences}, which constitutes private app-internal data storage that is off limits to other apps. Similarly, we store $DRID$s in private app-internal storage for each of the last 14 generations using the SHA-256 hash function. The \emph{seed} is only needed to generate $DRID$s, which we use to generate $RPID$s periodically.

We implement our random ID broadcast protocol as well as the data frame transmission and reception modules. To reduce audibility, our implementation uses ultrasonic waves from 17.5 kHz to 22.6 kHz. In particular, it supports two acoustic channels, each of which uses 54 frequencies. The interval between adjacent frequencies is 46.875 Hz. The first two groups of 21 frequencies are assigned to deliver the odd and even symbols discussed in \S\ref{sub_data_frame_chunk}, respectively. We reserve the  remaining frequencies to separate adjacent channels for reliable communications. 

\paragraph{The Server} The \sysname server only stores $DRID$s and corresponding dates for users who have tested positive for COVID-19. Mobile app users periodically retrieve these $DRID$s to determine whether contact with these users occurred. For efficiency, the server only pushes newly added data to the client. Also, the integrity of the server data is protected, and we require authentication in order for users who are diagnosed positive to upload their $DRID$s to the server.

\section{Evaluation}
\label{sec:evaluation}
In this section, we present \sysname's experimental results. First, we describe our experimental setup in \S\ref{sub:setup}. We discuss the results of effective tracing distance in \S\ref{eva_eft_tck_dis} and energy consumption in \S\ref{pgh_bat_noisy_in}. Finally, we compare \sysname with Bluetooth solutions in \S\ref{sub:bluetooth}.

\subsection{Experimental Setup}
\label{sub:setup}

We evaluate \sysname's ability to trace high-risk contact activities in two typical scenarios: \emph{Noisy In-Pocket} scenarios and \emph{Quiet Out-of-Pocket} scenarios, as environmental noise and occlusion are two major factors that affect \sysname's contact tracing quality. These two scenarios represent two extreme working scenarios of \sysname under the effects of noise and occlusion level. In other words, we believe that in the Quiet Out-of-Pocket scenario, \sysname receives the least interference, which would result in the longest tracing distance; on the other hand, in the Noisy In-Pocket scenario, \sysname experiences significant interference, which would result in its shortest tracing distance. We use the group tracing success rate (defined below) to evaluate the reliability of \sysname.
Suppose there are $n$ users in a group in physical proximity. For any user $u_{i}$, we denote the number of other users detected (received at least one $PRID$) by $u_{i}$ in 10 minutes as $k_{i}$. We define the \emph{individual tracing success rate} $R_{i} = \frac{k_{i}}{n-1}$, and the \emph{group tracing success rate} $r = \overline{R_{i}}$.

We also evaluate \sysname's energy consumption to determine how \sysname affects the battery life of mobile devices. We compare \sysname to Bluetooth Low Energy (BLE) in terms of its ability to permeate a wall in order to highlight \sysname's advantages in contact tracing against COVID-19. Throughout the evaluation, all mobile phones that run \sysname are Google Pixel 4s running Android 9+.

\subsection{Evaluation of Effective Tracing Distance}\label{eva_eft_tck_dis}
To evaluate the effective tracing distance, we place four Pixel 4 mobile phones at the vertices of a square on a flat hardwood floor. All phones face up while their speakers all point north. Each phone keeps broadcasting one \textit{RPID} at full volume for 10 minutes (the \textit{RPID} changes every 10 minutes) while the data frame that carries one \textit{RPID} is sent every 50 to 70 seconds. Each phone has a distinct \textit{RPID}. We collect the number of \textit{RPID}s received by individual phones every 10 minutes in order to calculate the group tracing success rate. We evaluate the group tracing success rate performance over the distance among phones (edges of the square). The experiments for each distance in each scenario are repeated ten times to reduce random error. The output of this evaluation are curves that indicate the group tracing success rate (average by times of experiments) at different distances for each scenario.

\begin{figure}[H]
    \centering
    \includegraphics[width=0.5\linewidth]{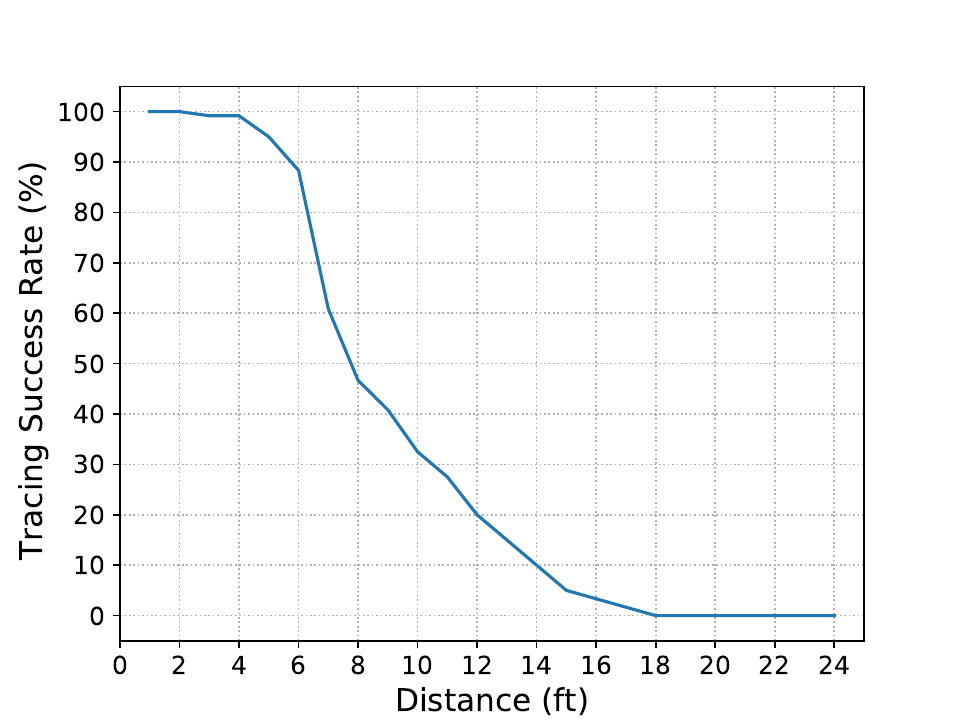}
    
    \caption{Tracing success rate of 4 mobile phones with different tracing distances under the Noisy In-Pocket scenario. The performance decreases when the distance between people exceeds 6 feet, suggesting \sysname's promise to reflect the natural properties of human contact.}
    \label{fig:exp_noisy_in}
\end{figure}

\paragraph{Noisy In-Pocket Scenario}\label{sub_noisy_in}
This scenario is common in universities or public transportation systems where the environmental noise is loud and everyone has a mobile phone in his/her pocket. In order to simulate the noisy environment, we place a mobile phone at the center of all other phones playing music randomly from Google Play Music at 52\% volume. To simulate the in-pocket scenario, each phone is completely covered by a thin fabric. \autoref{fig:exp_noisy_in} shows the success rate of tracing among 4 phones in the Noisy In-Pocket scenario. The first part of the curve (0 to 5 feet) consistently exceeds 95\%, which means that \sysname can trace contacts among people who spend more than 10 minutes within 5 feet of each other. The second part of the curve (6 to 12 feet) indicates that \sysname's success rate in contact tracking decreases rapidly when the distance between people is larger than 6 feet. The third part of the curve ($\ge$ 13 feet) shows that \sysname is unable to track people who are far from the phone even if their distance remains unchanged for 10 minutes. In conclusion, \sysname is a promising solution for contact tracing against COVID-19 in Noisy In-Pocket scenarios when phones broadcast \emph{PRID}s at full volume, as its effective tracing distance is close to 6 feet.

\begin{figure}[t]
    \centering
    \subfigure[Full Volume]{
        \includegraphics[width=0.45\linewidth]{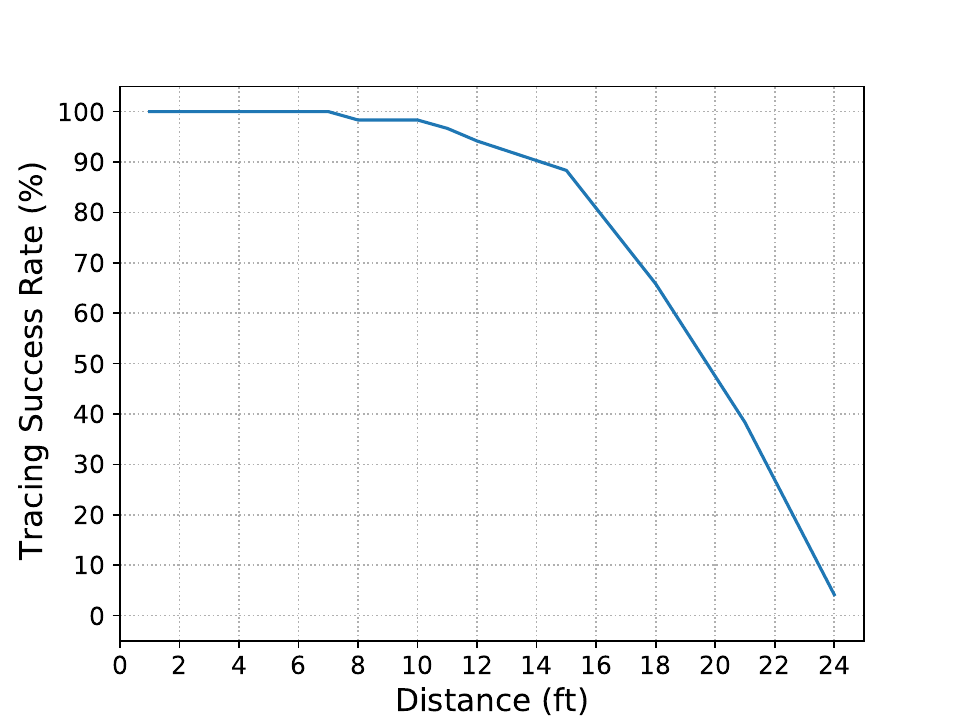}
        \label{fig:exp_quiet_out}
    }
    \subfigure[Non-full Volumes]{
        \includegraphics[width=0.45\linewidth]{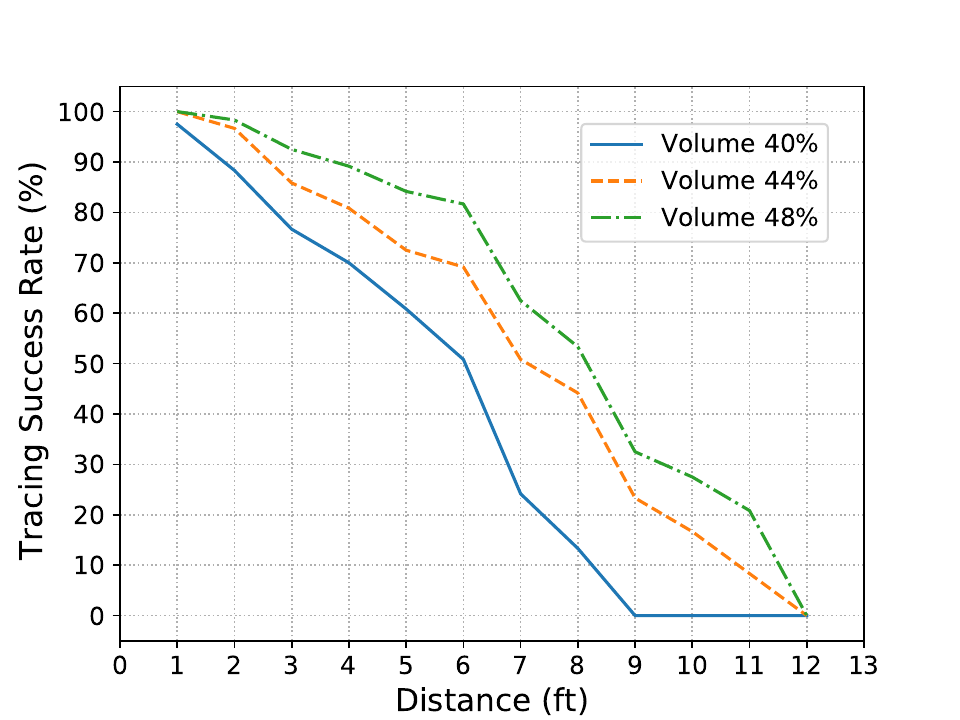}
        \label{fig:exp_quiet_vol}
    }
    \caption{Communication accuracy of \sysname under the Quiet Out-of-Pocket scenario. We show tracking success rates at different volume levels. By adjusting the volume level, \sysname  scales to various scenarios according to different effective tracing distances.}
    \label{fig:exp_quite_out_all}
\end{figure}

\paragraph{Quiet Out-of-Pocket Scenario}\label{sub_quiet_out}
This scenario corresponds to an airport departure gate (an hour before takeoff) or inside a library. We use a similar configuration as in \S\ref{eva_eft_tck_dis}: 4 phones are placed on a flat hardwood floor at vertices of a square. The difference is that the phone sending background noise and the fabric are removed for the Quiet Out-of-Pocket scenario. \autoref{fig:exp_quite_out_all}(a) shows the success rate of tracing between 4 Pixel 4 phones in the Quiet Out-of-Pocket scenario. The curve is constantly at 100\% until a distance of 12 feet, which means that in this scenario, \sysname can accurately track users within a 12-foot distance if this distance is fixed for at least 10 minutes. After 12 feet, the figure shows a rapid decrease of the success rate. We conclude that the effective tracing distance of \sysname in the Quiet Out-of-Pocket scenario is near 12 feet. \sysname is not a good fit for contact tracing against COVID-19 in the Quiet Out-of-Pocket scenario when devices broadcast at full volume, as its effective tracing distance greatly exceeds 6 feet.

In \autoref{fig:exp_quite_out_all}(a), the effective tracing range is up to 12 feet in the Quiet Out-of-Pocket scenario where phones broadcast at full volume. We evaluate the effective tracing distance under different volume levels, using the same setup with experiments in the Quiet Out-of-Pocket scenario but with varying volume levels. As there are 25 volume levels in the Pixel 4, we divide volume level between mute and full volume using 25 volume levels; hence, each pair of consecutive volume levels differs by 4\%. First, we adjust the volume level to determine the minimal volume level that allows \sysname to get a reasonable success rate at 6 feet (success rate $\ge$ 50\%). Next, we find the minimal volume level that allows \sysname to get a reliable success rate at 6 feet (success rate $\ge$ 90\%). Afterwards, we explore the tracing success rate curve with each volume level between minimal and maximal volume.  \autoref{fig:exp_quite_out_all}(b) shows that \sysname's tracking success rate using different volume levels in the Quiet Out-of-Pocket scenario. At 48\% volume level, the curve in the figure indicates that when all phones broadcast \textit{RPID}s, the turning point of tracking success rate is about 6 feet, which means that the effective tracking distance is at about 6 feet. By adjusting the volume level, it is possible to limit \sysname's effective tracking distance near 6 feet. However, since the two scenarios we involved in the evaluation have different ``best volume'' levels, the devices need to detect ambient environmental noise (in or out of pocket) to determine the volume level used to broadcast RPIDs. This is part of our future work.

\paragraph{Remark}\label{conclusion_effective_tracking_distance} With our evaluation in two scenarios, we conclude that \sysname is able to track high-risk contact made between users. We notice, however, that different scenarios require different volume levels to control the effective tracking distance. Therefore, \sysname needs to adjust the volume level according to the kind of scenario the devices are in, which forms part of our future work.
\begin{figure}
    
    \centering
    \begin{minipage}[t]{0.45\linewidth}
        \centering
        \includegraphics[width=0.98\linewidth]{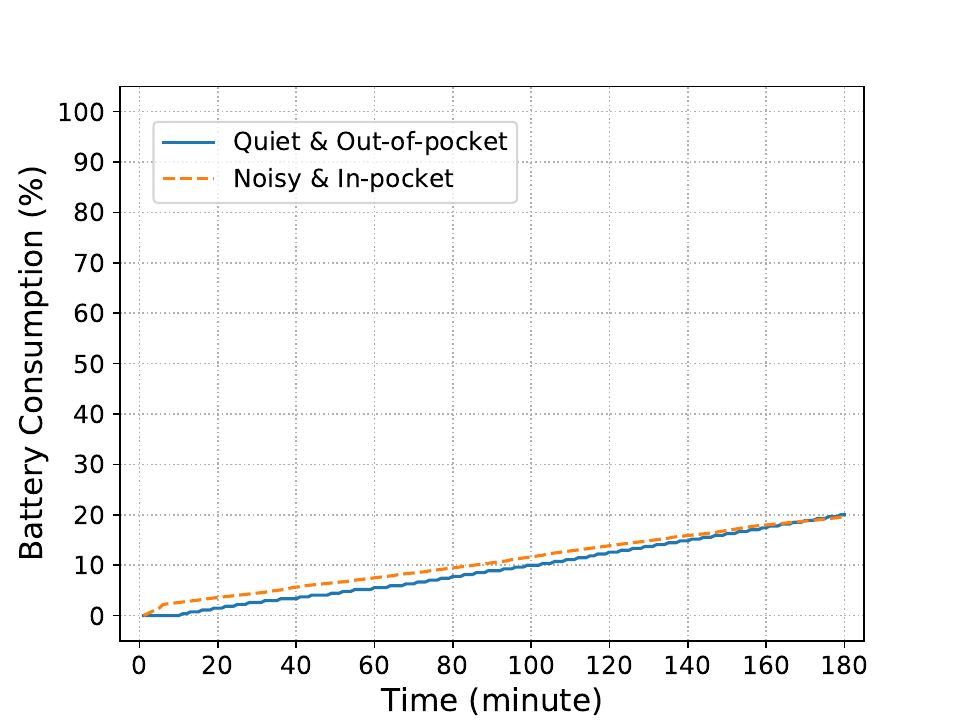}
        \caption{Power consumption of \sysname in different scenarios. The overall consumption is satisfactory for practical usage.}
        \label{fig:exp_power}
    \end{minipage}
    \quad
    \begin{minipage}[t]{0.45\linewidth}
        \centering
        \includegraphics[width=1\linewidth]{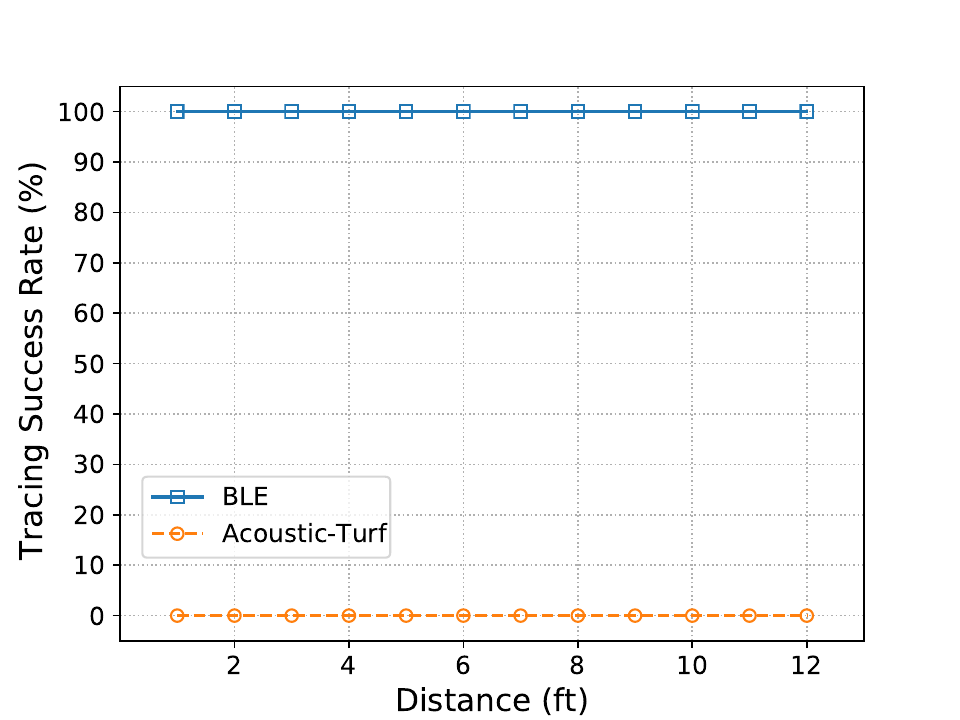}
        \caption{Success rate comparison between \sysname and BLE on through-wall communication.}
        \label{fig:exp_comparison}
    \end{minipage}
\end{figure}
\subsection{Evaluation of Energy Consumption}\label{pgh_bat_noisy_in}
To evaluate the battery use of \sysname in the two scenarios, we use the same configuration of devices as in \S\ref{eva_eft_tck_dis}. However, we fix the distance between Pixel 4 devices at exactly 6 feet and have all devices fully charged before start running \sysname. All devices keep broadcasting and receiving \textit{RPID} for three hours. We check the battery in the \sysname app every 60 seconds to construct \autoref{fig:exp_power}, which indicates that \sysname uses 6\% of battery each hour(on average) when both broadcasting and receiving \textit{RPID} in both scenarios at full volume. The battery consumption in the Noisy In-Pocket scenario is slightly higher than that in the Quiet Out-of-Pocket scenario. An explanation for this phenomenon is that in the Noisy In-Pocket scenario, \sysname increases the unit number per data frame to ensure the required tracking success rate, which slightly increases power consumption. Hence, we conclude that running \sysname in the background does not reduce battery lifetime to under 8 hours, which lets users charge their devices before their batteries are depleted.

\subsection{Comparison to Bluetooth-Based Solutions}
\label{sub:bluetooth}
A major concern about contact tracing using Bluetooth-based solutions is that such wireless signals can permeate through certain obstacles. This means that these solutions regard people separated by a wall as having been in contact with each other, whereas we know this is not the case.
In this section, we compare \sysname to BLE with a focus on the communication performance when individuals are separated by a wall.
Concretely, we put 2 Pixel 4 devices on each side of a wall on the hardwood floor without fabric covers. We control the distance between these devices to ensure they are placed at vertices of a square. Following \S\ref{eva_eft_tck_dis}, all devices start broadcast and receive \textit{RPID} at the same time for 10 minutes; we collect the number of \textit{RPID}s received by each device to evaluate \sysname's ability to permeate the wall. In addition, the devices run a BLE broadcasting app under the same setting.
\autoref{fig:exp_comparison} shows the success rates of \sysname and BLE. The figure shows that \sysname does not track people who are on different sides of a wall, which reflects the natural properties of contact tracing for COVID-19.

\section{Discussions and Future Work}
\label{sec:discussion}
\paragraph{Further Design on Acoustic Communications and Random ID Generation} Our current design for acoustic communication illustrates the feasibility of such communication on mobile phones for contact tracing spanning a 6-foot range among multiple users in occluded and noisy environments. Further improvements to our design are possible, including increased reliability and energy efficient communication for the same purpose. In addition, we can extend our design for other applications (such as IoT ones) by improving its throughput over various distances. Our current cryptographic protocol can also be improved in many ways. For instance, we can adopt a similar design as Apple and Google's for generating each day's $DRID$ independently instead of using a hash chain. To improve the precision of matching, we may restrict infected users to upload $RPID$s  during certain time windows only (as DP3T~\cite{DP3T} proposed) instead of regenerating $RPID$s from $DRID$s. In addition, our system does not keep IDs of infected users secret; we can leverage private set intersection~\cite{de2010practical} for greater privacy.

\paragraph{More Thorough Implementation and Evaluation} In the above sections, we reported our implementation and evaluation on Android mobile phones. Due to COVID-19, all authors have been staying at home while conducting this research. It is difficult for us to implement \sysname on multiple platforms (such as iOS) and conduct more thorough evaluations with various numbers of users in more practical working scenarios. 
We compare the tracing success rate of \sysname in Noisy In-Pocket and Quiet Out-of-Pocket scenarios in \S\ref{sub_noisy_in} with 4 mobile phones. The tracing success rate curve decreases at around 6 feet in the Noisy In-Pocket scenario. However, the curves of the Quiet Out-of-Pocket scenario decrease at different ranges depending on the speaker volumes. Similar phenomena occur in scenarios with fewer phones. Automatically adjusting various parameters based on working scenarios to ensure contact tracing ranges are around 6 feet remains an important part of our future research.

\paragraph{Other Improvements} We use ultrasonic frequencies ranging from 17,500\,Hz to 22,600\,Hz. However, due to BASK modulation, our system generates some audible noise, even using carrier waves above the audible frequency threshold. In future work, we aim to achieve inaudible acoustic communication in practice. Our system uses mobile phones' speakers and microphones to realize acoustic communication for contact tracing. Questions may arise about the impact of applications (e.g., phone calls and music playback) on speaker and microphone usage. Although our preliminary tests on this matter show limited impacts, this remains an area for future research. Acoustic signals offer the potential for sensing finer-grained contact detection. In addition, we can incorporate other sensors on mobile phones (such as compasses) in order to infer face orientations, handshakes, and conversations.

\section{Conclusion}
\label{sec:conclusion}
This paper proposed \sysname, a novel contact tracing system that preserves users' privacy by employing acoustic signals on commercial off-the-shelf (COTS) Android phones. \sysname performs automated contact tracing and enables infected users to upload their random identifiers (at their discretion) to public health authorities. \sysname realized such functionality using a privacy-preserving cryptographic protocol for security and an acoustic communications protocol. We have implemented \sysname on COTS mobile phones. Our experimental evaluation showed the promise of our system for privacy-preserving contact tracing using acoustic signals. 
In particular, \sysname exhibited better performance than Bluetooth-based contact tracing solutions in certain scenarios (such as people on different sides of walls). In our system, there is a clear ``turf'' effect where contact tracing rates dramatically decrease at certain distances. For instance, with 4 phones and the Noisy In-Pocket working scenario, the tracing rate decreased rapidly beyond a 6-foot range. There are multiple avenues to improve \sysname, and our immediate future work is to extend \sysname to mobile devices running iOS and evaluate the system's performance on an ensemble of devices running both iOS and Android operating systems.

\section*{Acknowledgement}
The work was supported in part by the National Science Foundation (NSF) under Grant No. CNS 2028547. Any opinions, findings, conclusions, and recommendations in this paper are those of the authors and do not necessarily reflect the views of the funding agencies.

\bibliographystyle{ACM-Reference-Format}
\bibliography{acmart}


\begin{thebibliography}{28}


\ifx \showCODEN    \undefined \def \showCODEN     #1{\unskip}     \fi
\ifx \showDOI      \undefined \def \showDOI       #1{#1}\fi
\ifx \showISBNx    \undefined \def \showISBNx     #1{\unskip}     \fi
\ifx \showISBNxiii \undefined \def \showISBNxiii  #1{\unskip}     \fi
\ifx \showISSN     \undefined \def \showISSN      #1{\unskip}     \fi
\ifx \showLCCN     \undefined \def \showLCCN      #1{\unskip}     \fi
\ifx \shownote     \undefined \def \shownote      #1{#1}          \fi
\ifx \showarticletitle \undefined \def \showarticletitle #1{#1}   \fi
\ifx \showURL      \undefined \def \showURL       {\relax}        \fi
\providecommand\bibfield[2]{#2}
\providecommand\bibinfo[2]{#2}
\providecommand\natexlab[1]{#1}
\providecommand\showeprint[2][]{arXiv:#2}

\bibitem[\protect\citeauthoryear{Al~Qathrady and Helmy}{Al~Qathrady and
  Helmy}{2017}]%
        {al2017improving}
\bibfield{author}{\bibinfo{person}{Mimonah Al~Qathrady} {and}
  \bibinfo{person}{Ahmed Helmy}.} \bibinfo{year}{2017}\natexlab{}.
\newblock \showarticletitle{Improving BLE Distance Estimation and
  Classification Using TX Power and Machine Learning: A Comparative Analysis}.
  In \bibinfo{booktitle}{\emph{Proceedings of the 20th ACM International
  Conference on Modelling, Analysis and Simulation of Wireless and Mobile
  Systems}}. \bibinfo{publisher}{ACM}, \bibinfo{pages}{79--83}.
\newblock


\bibitem[\protect\citeauthoryear{Apple and Google}{Apple and Google}{2020}]%
        {AG:PPCT}
\bibfield{author}{\bibinfo{person}{Privacy-Preserving Contact~Tracing Apple}
  {and} \bibinfo{person}{Google}.} \bibinfo{year}{2020}\natexlab{}.
\newblock
  \bibinfo{howpublished}{\url{https://www.apple.com/covid19/contacttracing/}}.
\newblock
\newblock
\shownote{(Accessed on 04/23/2020).}


\bibitem[\protect\citeauthoryear{Bostock}{Bostock}{2020}]%
        {Expertsw13:online}
\bibfield{author}{\bibinfo{person}{Bill Bostock}.}
  \bibinfo{year}{2020}\natexlab{}.
\newblock \bibinfo{title}{Experts wary of rushed COVID-19 vaccine after Fauci
  says 12-18 months - Business Insider}.
\newblock
  \bibinfo{howpublished}{\url{https://www.businessinsider.com/coronavirus-vaccine-quest-18-months-fauci-experts-flag-dangers-testing-2020-4}}.
\newblock
\newblock
\shownote{(Accessed on 04/23/2020).}


\bibitem[\protect\citeauthoryear{Canetti, Trachtenberg, and Varia}{Canetti
  et~al\mbox{.}}{2020}]%
        {Canetti2020AnonymousCD}
\bibfield{author}{\bibinfo{person}{Ran Canetti}, \bibinfo{person}{Ari
  Trachtenberg}, {and} \bibinfo{person}{Mayank Varia}.}
  \bibinfo{year}{2020}\natexlab{}.
\newblock \showarticletitle{Anonymous Collocation Discovery: Harnessing Privacy
  to Tame the Coronavirus}.
\newblock \bibinfo{journal}{\emph{arXiv: Computers and Society}}
  (\bibinfo{year}{2020}).
\newblock


\bibitem[\protect\citeauthoryear{Chan, Foster, Gollakota, Horvitz, Jaeger,
  Kakade, Kohno, Langford, Larson, Singanamalla, Sunshine, and Tessaro}{Chan
  et~al\mbox{.}}{2020}]%
        {Chan2020PACTPS}
\bibfield{author}{\bibinfo{person}{Justin Chan}, \bibinfo{person}{Dean~P.
  Foster}, \bibinfo{person}{Shyam Gollakota}, \bibinfo{person}{Eric Horvitz},
  \bibinfo{person}{Joseph Jaeger}, \bibinfo{person}{Sham~M. Kakade},
  \bibinfo{person}{Tadayoshi Kohno}, \bibinfo{person}{John Langford},
  \bibinfo{person}{Jonathan Larson}, \bibinfo{person}{Sudheesh Singanamalla},
  \bibinfo{person}{Jacob~E. Sunshine}, {and} \bibinfo{person}{Stefano
  Tessaro}.} \bibinfo{year}{2020}\natexlab{}.
\newblock \showarticletitle{PACT: Privacy Sensitive Protocols and Mechanisms
  for Mobile Contact Tracing}.
\newblock \bibinfo{journal}{\emph{ArXiv}}  \bibinfo{volume}{abs/2004.03544}
  (\bibinfo{year}{2020}).
\newblock


\bibitem[\protect\citeauthoryear{Community}{Community}{2020}]%
        {opentrac5:online}
\bibfield{author}{\bibinfo{person}{OpenTrace Community}.}
  \bibinfo{year}{2020}\natexlab{}.
\newblock \bibinfo{title}{OpenTrace Calibration. Device calibration data and
  Trial Methodologies for testing implementations of the BlueTrace protocol.}
\newblock
  \bibinfo{howpublished}{\url{https://github.com/opentrace-community/opentrace-calibration}}.
\newblock
\newblock
\shownote{(Accessed on 05/11/2020).}


\bibitem[\protect\citeauthoryear{CovidSafe}{CovidSafe}{2020}]%
        {CovidSaf66}
\bibfield{author}{\bibinfo{person}{CovidSafe}.}
  \bibinfo{year}{2020}\natexlab{}.
\newblock \bibinfo{howpublished}{\url{https://github.com/CovidSafe}}.
\newblock
\newblock
\shownote{(Accessed on 04/23/2020).}


\bibitem[\protect\citeauthoryear{Crocker, Opsahl, and Cyphers}{Crocker
  et~al\mbox{.}}{2020}]%
        {Bluetoot:Challenges}
\bibfield{author}{\bibinfo{person}{Andrew Crocker}, \bibinfo{person}{Kurt
  Opsahl}, {and} \bibinfo{person}{Bennett Cyphers}.}
  \bibinfo{year}{2020}\natexlab{}.
\newblock \bibinfo{title}{Bluetooth contact tracing needs bigger, better data}.
\newblock
  \bibinfo{howpublished}{\url{https://www.technologyreview.com/2020/04/22/1000353/bluetooth-contact-tracing-needs-bigger-better-data/}}.
\newblock
\newblock
\shownote{(Accessed on 04/23/2020).}


\bibitem[\protect\citeauthoryear{Dave}{Dave}{2020}]%
        {Explaine84:online}
\bibfield{author}{\bibinfo{person}{Paresh Dave}.}
  \bibinfo{year}{2020}\natexlab{}.
\newblock \bibinfo{title}{Explainer: How smartphone apps can help 'contact
  trace' the new coronavirus - Reuters}.
\newblock
  \bibinfo{howpublished}{\url{https://www.reuters.com/article/us-health-coronavirus-tracing-apps-expla/explainer-how-smartphone-apps-can-help-contact-trace-the-new-coronavirus-idUSKCN21W2I8}}.
\newblock
\newblock
\shownote{(Accessed on 04/23/2020).}


\bibitem[\protect\citeauthoryear{De~Cristofaro and Tsudik}{De~Cristofaro and
  Tsudik}{2010}]%
        {de2010practical}
\bibfield{author}{\bibinfo{person}{Emiliano De~Cristofaro} {and}
  \bibinfo{person}{Gene Tsudik}.} \bibinfo{year}{2010}\natexlab{}.
\newblock \showarticletitle{Practical private set intersection protocols with
  linear complexity}. In \bibinfo{booktitle}{\emph{International Conference on
  Financial Cryptography and Data Security}}. Springer,
  \bibinfo{publisher}{Springer}, \bibinfo{pages}{143--159}.
\newblock


\bibitem[\protect\citeauthoryear{Fussell and Knight}{Fussell and
  Knight}{2020}]%
        {TheApple44:online}
\bibfield{author}{\bibinfo{person}{Sidney Fussell} {and} \bibinfo{person}{Will
  Knight}.} \bibinfo{year}{2020}\natexlab{}.
\newblock \bibinfo{title}{The Apple-Google Contact Tracing Plan Won't Stop
  Covid Alone}.
\newblock
  \bibinfo{howpublished}{\url{https://www.wired.com/story/apple-google-contact-tracing-wont-stop-covid-alone/}}.
\newblock
\newblock
\shownote{(Accessed on 04/23/2020).}


\bibitem[\protect\citeauthoryear{Loh}{Loh}{[n.d.]}]%
        {NOVID41:online}
\bibfield{author}{\bibinfo{person}{Po-Shen Loh}.}
  \bibinfo{year}{[n.d.]}\natexlab{}.
\newblock \bibinfo{title}{NOVID}.
\newblock \bibinfo{howpublished}{\url{https://www.novid.org/}}.
\newblock
\newblock
\shownote{(Accessed on 06/22/2020).}


\bibitem[\protect\citeauthoryear{Maccari and Cagno}{Maccari and Cagno}{2020}]%
        {Maccari2020DoWN}
\bibfield{author}{\bibinfo{person}{Leonardo Maccari} {and}
  \bibinfo{person}{Valeria Cagno}.} \bibinfo{year}{2020}\natexlab{}.
\newblock \showarticletitle{Do we need a Contact Tracing App?}
\newblock \bibinfo{journal}{\emph{ArXiv}}  \bibinfo{volume}{abs/2005.10187}
  (\bibinfo{year}{2020}).
\newblock


\bibitem[\protect\citeauthoryear{Nandakumar, Chintalapudi, Padmanabhan, and
  Venkatesan}{Nandakumar et~al\mbox{.}}{2013}]%
        {nandakumar2013dhwani}
\bibfield{author}{\bibinfo{person}{Rajalakshmi Nandakumar},
  \bibinfo{person}{Krishna~Kant Chintalapudi}, \bibinfo{person}{Venkat
  Padmanabhan}, {and} \bibinfo{person}{Ramarathnam Venkatesan}.}
  \bibinfo{year}{2013}\natexlab{}.
\newblock \showarticletitle{Dhwani: secure peer-to-peer acoustic NFC}.
\newblock \bibinfo{journal}{\emph{ACM SIGCOMM Computer Communication Review}}
  \bibinfo{volume}{43}, \bibinfo{number}{4} (\bibinfo{year}{2013}),
  \bibinfo{pages}{63--74}.
\newblock


\bibitem[\protect\citeauthoryear{Niyogi, Petrie, Leibrand, Gallagher, Hamish,
  Szabo, Danezis, Miers, de~Valence, and Reusche}{Niyogi et~al\mbox{.}}{2020}]%
        {TCN}
\bibfield{author}{\bibinfo{person}{Sourabh Niyogi}, \bibinfo{person}{James
  Petrie}, \bibinfo{person}{Scott Leibrand}, \bibinfo{person}{Jack Gallagher},
  \bibinfo{person}{Manu~Eder Hamish}, \bibinfo{person}{Zsombor Szabo},
  \bibinfo{person}{George Danezis}, \bibinfo{person}{Ian Miers},
  \bibinfo{person}{Henry de Valence}, {and} \bibinfo{person}{Daniel Reusche}.}
  \bibinfo{year}{2020}\natexlab{}.
\newblock \bibinfo{title}{TCNCoalition/TCN: Specification and reference
  implementation of the TCN Protocol for decentralized, privacy-preserving
  contact tracing.}
\newblock \bibinfo{howpublished}{\url{https://github.com/TCNCoalition/TCN}}.
\newblock
\newblock
\shownote{(Accessed on 04/23/2020).}


\bibitem[\protect\citeauthoryear{Normile}{Normile}{2020}]%
        {Coronavi89:online}
\bibfield{author}{\bibinfo{person}{Dennis Normile}.}
  \bibinfo{year}{2020}\natexlab{}.
\newblock \bibinfo{title}{Coronavirus cases have dropped sharply in South
  Korea. What’s the secret to its success?}
\newblock
  \bibinfo{howpublished}{\url{https://www.sciencemag.org/news/2020/03/coronavirus-cases-have-dropped-sharply-south-korea-whats-secret-its-success}}.
\newblock
\newblock
\shownote{(Accessed on 04/23/2020).}


\bibitem[\protect\citeauthoryear{Novak, Tang, and Li}{Novak
  et~al\mbox{.}}{2018}]%
        {novak2018ultrasound}
\bibfield{author}{\bibinfo{person}{Ed Novak}, \bibinfo{person}{Zhuofan Tang},
  {and} \bibinfo{person}{Qun Li}.} \bibinfo{year}{2018}\natexlab{}.
\newblock \showarticletitle{Ultrasound proximity networking on smart mobile
  devices for IoT applications}.
\newblock \bibinfo{journal}{\emph{IEEE Internet of Things Journal}}
  \bibinfo{volume}{6}, \bibinfo{number}{1} (\bibinfo{year}{2018}),
  \bibinfo{pages}{399--409}.
\newblock


\bibitem[\protect\citeauthoryear{of~Singapore}{of~Singapore}{2020}]%
        {TraceTogether}
\bibfield{author}{\bibinfo{person}{Government of Singapore}.}
  \bibinfo{year}{2020}\natexlab{}.
\newblock \bibinfo{title}{Trace Together, safer together}.
\newblock \bibinfo{howpublished}{\url{https://www.tracetogether.gov.sg}}.
\newblock
\newblock
\shownote{(Accessed on 04/23/2020).}


\bibitem[\protect\citeauthoryear{Project}{Project}{2019a}]%
        {Advertis93:online}
\bibfield{author}{\bibinfo{person}{Android Open~Source Project}.}
  \bibinfo{year}{2019}\natexlab{a}.
\newblock \bibinfo{title}{AdvertiseSettings.Builder}.
\newblock
  \bibinfo{howpublished}{\url{https://developer.android.com/reference/android/bluetooth/le/AdvertiseSettings.Builder\#setTxPowerLevel(int)}}.
\newblock
\newblock
\shownote{(Accessed on 05/11/2020).}


\bibitem[\protect\citeauthoryear{Project}{Project}{2019b}]%
        {ScanReco97:online}
\bibfield{author}{\bibinfo{person}{Android Open~Source Project}.}
  \bibinfo{year}{2019}\natexlab{b}.
\newblock \bibinfo{title}{ScanRecord}.
\newblock
  \bibinfo{howpublished}{\url{https://developer.android.com/reference/android/bluetooth/le/ScanRecord\#getTxPowerLevel()}}.
\newblock
\newblock
\shownote{(Accessed on 05/11/2020).}


\bibitem[\protect\citeauthoryear{Project}{Project}{2019c}]%
        {ScanResu12:online}
\bibfield{author}{\bibinfo{person}{Android Open~Source Project}.}
  \bibinfo{year}{2019}\natexlab{c}.
\newblock \bibinfo{title}{ScanResult}.
\newblock
  \bibinfo{howpublished}{\url{https://developer.android.com/reference/android/bluetooth/le/ScanResult\#getRssi()}}.
\newblock
\newblock
\shownote{(Accessed on 05/11/2020).}


\bibitem[\protect\citeauthoryear{Rivest, Callas, Canetti, Esvelt, Gillmor,
  Kalai, Lysyanskaya, Norige, Raskar, Shamir, Shen, Soibelman, Specter, Teague,
  Trachtenberg, Varia, Viera, Weitzner, Wilkinson, and Zissman}{Rivest
  et~al\mbox{.}}{2020}]%
        {PACT}
\bibfield{author}{\bibinfo{person}{Ronald~L. Rivest}, \bibinfo{person}{Jon
  Callas}, \bibinfo{person}{Ran Canetti}, \bibinfo{person}{Kevin Esvelt},
  \bibinfo{person}{Daniel~Kahn Gillmor}, \bibinfo{person}{Yael~Tauman Kalai},
  \bibinfo{person}{Anna Lysyanskaya}, \bibinfo{person}{Adam Norige},
  \bibinfo{person}{Ramesh Raskar}, \bibinfo{person}{Adi Shamir},
  \bibinfo{person}{Emily Shen}, \bibinfo{person}{Israel Soibelman},
  \bibinfo{person}{Michael Specter}, \bibinfo{person}{Vanessa Teague},
  \bibinfo{person}{Ari Trachtenberg}, \bibinfo{person}{Mayank Varia},
  \bibinfo{person}{Marc Viera}, \bibinfo{person}{Daniel Weitzner},
  \bibinfo{person}{John Wilkinson}, {and} \bibinfo{person}{Marc Zissman}.}
  \bibinfo{year}{2020}\natexlab{}.
\newblock \bibinfo{title}{The PACT protocol specification}.
\newblock
  \bibinfo{howpublished}{\url{https://pact.mit.edu/wp-content/uploads/2020/04/The-PACT-protocol-specification-ver-0.1.pdf}}.
\newblock
\newblock
\shownote{(Accessed on 04/23/2020).}


\bibitem[\protect\citeauthoryear{ROBust and privacy-presERving~proximity
  Tracing~protocol}{ROBust and privacy-presERving~proximity
  Tracing~protocol}{2020}]%
        {ROBERTRO82:online}
\bibfield{author}{\bibinfo{person}{ROBERT ROBust} {and}
  \bibinfo{person}{privacy-presERving~proximity Tracing~protocol}.}
  \bibinfo{year}{2020}\natexlab{}.
\newblock
  \bibinfo{howpublished}{\url{https://github.com/ROBERT-proximity-tracing}}.
\newblock
\newblock
\shownote{(Accessed on 05/12/2020).}


\bibitem[\protect\citeauthoryear{{Santagati} and {Melodia}}{{Santagati} and
  {Melodia}}{2017}]%
        {santagati2015u}
\bibfield{author}{\bibinfo{person}{G.~E. {Santagati}} {and} \bibinfo{person}{T.
  {Melodia}}.} \bibinfo{year}{2017}\natexlab{}.
\newblock \showarticletitle{A Software-Defined Ultrasonic Networking Framework
  for Wearable Devices}.
\newblock \bibinfo{journal}{\emph{IEEE/ACM Transactions on Networking}}
  \bibinfo{volume}{25}, \bibinfo{number}{2}, \bibinfo{pages}{960--973}.
\newblock


\bibitem[\protect\citeauthoryear{Troncoso, Payer, Hubaux, Salath{\'e}, Larus,
  Bugnion, Lueks, Stadler, Pyrgelis, Antonioli, et~al\mbox{.}}{Troncoso
  et~al\mbox{.}}{2020}]%
        {DP3T}
\bibfield{author}{\bibinfo{person}{Carmela Troncoso}, \bibinfo{person}{Mathias
  Payer}, \bibinfo{person}{Jean-Pierre Hubaux}, \bibinfo{person}{Marcel
  Salath{\'e}}, \bibinfo{person}{James Larus}, \bibinfo{person}{Edouard
  Bugnion}, \bibinfo{person}{Wouter Lueks}, \bibinfo{person}{Theresa Stadler},
  \bibinfo{person}{Apostolos Pyrgelis}, \bibinfo{person}{Daniele Antonioli},
  {et~al\mbox{.}}} \bibinfo{year}{2020}\natexlab{}.
\newblock \bibinfo{title}{Decentralized Privacy-Preserving Proximity Tracing}.
\newblock \bibinfo{howpublished}{\url{https://github.com/DP3T/documents}}.
\newblock
\newblock
\shownote{(Accessed on 04/23/2020).}


\bibitem[\protect\citeauthoryear{White, Fenwick, Becker-Mayer, Petrie, Szabo,
  Blank, Colligan, Hittle, Ingle, Nash, Nguyen, Schwaber, Veeraghanta,
  Voloshin, Arx, and Xue}{White et~al\mbox{.}}{2020}]%
        {CovidWatch}
\bibfield{author}{\bibinfo{person}{Tina White}, \bibinfo{person}{Rhys Fenwick},
  \bibinfo{person}{Isaiah Becker-Mayer}, \bibinfo{person}{James Petrie},
  \bibinfo{person}{Zsombor Szabo}, \bibinfo{person}{Daniel Blank},
  \bibinfo{person}{Jesse Colligan}, \bibinfo{person}{Mike Hittle},
  \bibinfo{person}{Mark Ingle}, \bibinfo{person}{Oliver Nash},
  \bibinfo{person}{Victoria Nguyen}, \bibinfo{person}{Jeff Schwaber},
  \bibinfo{person}{Akhil Veeraghanta}, \bibinfo{person}{Mikhail Voloshin},
  \bibinfo{person}{Sydney~Von Arx}, {and} \bibinfo{person}{Helen Xue}.}
  \bibinfo{year}{2020}\natexlab{}.
\newblock \bibinfo{title}{Slowing the spread of infectious diseases using
  crowdsourced data}.
\newblock \bibinfo{howpublished}{\url{https://www.covid-watch.org/article}}.
\newblock
\newblock
\shownote{(Accessed on 04/23/2020).}


\bibitem[\protect\citeauthoryear{Zafari, Gkelias, and Leung}{Zafari
  et~al\mbox{.}}{2019}]%
        {zafari2019survey}
\bibfield{author}{\bibinfo{person}{Faheem Zafari}, \bibinfo{person}{Athanasios
  Gkelias}, {and} \bibinfo{person}{Kin~K Leung}.}
  \bibinfo{year}{2019}\natexlab{}.
\newblock \showarticletitle{A survey of indoor localization systems and
  technologies}.
\newblock \bibinfo{journal}{\emph{IEEE Communications Surveys \& Tutorials}}
  \bibinfo{volume}{21}, \bibinfo{number}{3} (\bibinfo{year}{2019}),
  \bibinfo{pages}{2568--2599}.
\newblock


\bibitem[\protect\citeauthoryear{Zhang, Du, Zhou, Li, and Mohapatra}{Zhang
  et~al\mbox{.}}{2017}]%
        {zhang2017acoustic}
\bibfield{author}{\bibinfo{person}{Huanle Zhang}, \bibinfo{person}{Wan Du},
  \bibinfo{person}{Pengfei Zhou}, \bibinfo{person}{Mo Li}, {and}
  \bibinfo{person}{Prasant Mohapatra}.} \bibinfo{year}{2017}\natexlab{}.
\newblock \showarticletitle{An acoustic-based encounter profiling system}.
\newblock \bibinfo{journal}{\emph{IEEE Transactions on Mobile Computing}}
  \bibinfo{volume}{17}, \bibinfo{number}{8} (\bibinfo{year}{2017}),
  \bibinfo{pages}{1750--1763}.
\newblock


\end{thebibliography}

\end{document}